# ViewfinderVR: Configurable Viewfinder for Selection of Distant Objects in VR

Woojoo Kim, Shuping Xiong[*]

Department of Industrial and Systems Engineering, College of Engineering, Korea Advanced Institute of Science and Technology (KAIST), 291 Daehak-ro, Yuseong-gu, Daejeon 34141, Republic of Korea; {xml1324, shupingx}@kaist.ac.kr

* Corresponding author (Prof. Shuping Xiong), Telephone: +82-42-350-3132, Fax: +82-42-350-3110, Email: shupingx@kaist.ac.kr

**Abstract.** Selection is one of the fundamental user interactions in virtual reality (VR) and 3D user interaction, and raycasting has been one of the most popular object selection techniques in VR. However, the selection of small or distant objects through raycasting has been known to be difficult. To overcome this limitation, this study proposed a new technique called ViewfinderVR for improved selection of distant objects in VR, utilizing a virtual viewfinder panel with a modern adaptation of the through-the-lens metaphor. ViewfinderVR enables faster and more accurate target selection by allowing customization of the interaction space projected onto a virtual panel within reach, and users can select objects reflected on the panel with either ray-based or touch interaction. Experimental results of Fitts' law-based tests with 20 participants showed that ViewfinderVR outperformed traditional raycasting in terms of task performance (movement time, error rate, and throughput) and perceived workload (NASA-TLX ratings), where touch interaction was superior to ray-based interaction. The associated user behavior was also recorded and analyzed to understand the underlying reasons for the improved task performance and reduced workload. The proposed technique can be used in VR applications to enhance the selection of distant objects.

## 1. Introduction

Selection is one of the fundamental user interactions in virtual reality (VR) and 3D user interfaces (LaViola Jr. et al. 2017). As the selection is typically a prerequisite to manipulation of virtual objects, inferior selection techniques can easily harm the usability of VR interaction in general. Selection techniques can largely be divided into two categories: virtual hand and virtual pointing (Argelaguet and Andujar 2013). Virtual hand techniques allow users to select objects through the use of a virtual hand (Poupyrev et al. 1998), while virtual pointing techniques (also known as raycasting) select objects by pointing with a virtual ray (Liang and Green 1994; Mine 1995).

Although the simple virtual hand technique can enable more natural and better-performing interaction with isomorph mapping of the virtual and real hands hence superimposing visual and motor spaces (Mine et al. 1997; Wang and Mackenzie 1999; Lemmerman and LaViola Jr. 2007), virtual pointing techniques have become one of the most popular 3D selection techniques due to its simplicity and generality (Bowman and Hodges 1997; Lee et al. 2003; Andujar and Argelaguet 2007; Bacim et al. 2013), allowing users to select out-of-reach objects with relatively less physical movement. Moreover, numerous user studies have reported that raycasting results in better selection performance compared to competing 3D selection techniques (LaViola Jr. et al. 2017) including the hand extension approaches (Grossman and Balakrishnan 2006; Vanacken et al. 2007).

However, the selection of small or distant objects via raycasting has been reported to be difficult. A number of studies have found that raycasting is sensitive to natural hand tremor at far distances thus yields high error rates when selecting smaller targets (Herndon et al. 1994; Steed and Parker 2004; Cockburn et al. 2011; Lu et al. 2020; Brasier et al. 2020), as the selection occurs in mid-air while no physical support is given to hands (Lindeman et al. 1999). The Heisenberg effect (Bowman et al. 2001), indicating a change in the tool orientation happens by the action of selection trigger (e.g. a button press), is also known to induce target misses, accounting for near 30% of the overall errors in a pointing task (Wolf et al. 2020). Furthermore, positional and rotational jitter from the tracking device can negatively impact the selection accuracy (Weise et al. 2020; Batmaz et al. 2021). This could be especially problematic when vision-tracked freehand input is used where the tracking noise becomes substantial. It has been suggested to use the secondary position on the body to anchor the ray's direction to stabilize the ray in such a case (Facebook 2021), similarly to the fixed-origin pointing introduced in the work of Jota et al. (2010).

Many studies have attempted to enhance virtual pointing techniques via various approaches. One approach to resolve the precision issue of the vector-based pointing which relied on a single vector of the virtual ray was to introduce a volume such as a cone instead (Liang and Green 1994; Forsberg et al. 1996), although it required disambiguation when the volume contained multiple objects. Disambiguation could be done progressively by allowing users to specify the target from the priorly selected potential targets arranged in a menu (Kopper et al. 2011; Cashion et al. 2012) or to control a depth cursor (Grossman and Balakrishnan 2006; Baloup et al. 2019). It could also be done heuristically by assigning scores to potential targets and select the target with the highest score. Techniques used angle offset or distance to the original ray to assign higher scores to the nearest object (Wingrave et al. 2002; Grossman and Balakrishnan 2005; Steinicke et al. 2006; Vanacken et al. 2007; Cashion et al. 2013; Lu et al. 2020), while some more complex examples additionally employed a queue of scores accumulated over multiple timeframes (De Haan et al. 2005; Ortega 2013; Moore et al. 2018). The other approach to improve virtual pointing was to dynamically modify the control-display ratio when more precision is needed. Alteration of the control-display ratio could be done either manually (Vogel and Balakrishnan 2005; Kopper et al. 2010) or automatically based on the user's hand velocity (Frees et al. 2007; König et al. 2009; Gallo et al. 2010).

Meanwhile, some techniques have used an indirect approach where the user manipulates virtual objects by interacting with their copied representation. *World-in-Miniature* (Stoakley et al. 1995) was one of the first few techniques which introduced the use of proxies for object manipulation, further enhanced with an addition of the option to scale the proxies to a convenient size for interaction (Pierce et al. 1999; Wingrave et al. 2006). Indirect proxy techniques are known to benefit users by not only providing a separate exocentric view of the world which allows users to observe the world with different angles and distances thereby better understand the arrangements of objects (Wingrave et al. 2006), but also enabling interaction directly through hands hence avoiding body distortion and negative impact on body ownership (Bowman et al. 1997). A more advanced and comprehensive example of this approach that encompasses interaction with proxies of spaces instead of objects can be found in the work of Pohl et al. (2021).

Studies have shown 2D selection can outperform 3D selection (Pierce et al. 1997; Ware and Lowther 1997; Natapov and MacKenzie 2010; Teather and Stuerzlinger 2013; Ramcharitar and Teather 2018), thus corresponding adaptation can be made to benefit interaction with proxies. The *Through-the-lens* metaphor (Stoev and Schmalstieg 2002) is an early example of leveraging the benefits of 2D interaction in 3D proxies with the use of an additional viewport of the virtual world in a form of a floating virtual panel. Similar adaptations were presented by several studies since. Argelaguet and Andujar (2009) locally flattened targets onto a small virtual screen in stereoscopic displays to avoid binocular parallax. Surale et al. (2019) explored the use of a tangible viewport attached to a physical tablet for selecting distant VR objects by tapping on the tablet screen. Li et al. (2021) proposed the viewport to reflect the virtual world like a mirror for the selection of occluded and distanced objects in VR.

Despite the promising prospect of such a metaphor, only limited studies have tested its usability on target selection throughout the user study. Clergeaud and Guitton (2017) implemented the *Through-the-lens* interaction technique based on a cylindrical window displaying a 360° panoramic image for the object finding task, yet its full potential with the advantages from the simple virtual hand technique and optimization of the interaction space is undiscovered. In this study, we propose the *ViewfinderVR* technique, a virtual viewfinder panel with a modern adaptation of the *Through-the-lens* metaphor for a way to better select distant objects inside the virtual environment. ViewfinderVR enables faster and more accurate target selection by allowing customization of the interaction space projected onto a virtual panel within reach. We conducted a user study to measure its performance against the raycasting technique in a Fitts' Law-based test to understand the benefits and limitations of the proposed technique.

## 2. Methods

### 2.1. Fitts' Law

Fitts' law (Fitts 1954) is a predictive model that characterizes kinematics of human psychomotor behavior on the pointing and selection task. As it was proved that Fitts' law holds for the arm kinematics in 3D trajectories (MacKenzie et al. 1987), it has been widely used to evaluate the performance of selection techniques and pointing devices of 3D selection tasks (Steed 2006; Kopper et al. 2010; Teather and Stuerzlinger 2011; Ramcharitar and Teather 2018; Li et al. 2018). The model demonstrates the linear relationship between the index of difficulty (*ID*) of the task and movement time (*MT*), given as:

$$MT = a + b \times ID \quad (1)$$

where

$$ID = log_2 \left(\frac{D}{W} + 1\right) \quad (2)$$

*D* is the target distance and *W* is the target size (Fig. 1), whereas *a* and *b* are parameters empirically derived by linear regression. Fitts' law has been utilized as a tool for evaluating pointing devices through ISO 9241-9 (ISO 2002). The standard suggests using the multi-directional tapping task that covers different movement directions by arranging the targets in a circular pattern (Fig. 1). In addition, it recommends the use of throughput (*TP*) that quantifies the information processing rate of the human by combining both speed and accuracy (Soukoreff and MacKenzie 2004), calculated as:

$$TP = \frac{ID_e}{MT} \quad (3)$$

where

$$ID_e = log_2 \left(\frac{D_e}{W_e} + 1\right) \quad (4)$$

and

$$W_e = 4.133 \times SD_x \quad (5)$$

*D* and *W* are adjusted based on the user's actual movement trajectory to represent the effective target distance ($D_e$) and the effective target size ($W_e$). $D_e$ is calculated by averaging movement distance, while $W_e$ is calculated by estimating $SD_x$, the standard deviation of lengths relative to the target center projected onto the task axis. $SD_x$ is then multiplied by 4.133 to represent 96% of the target hit coordinates (Z-score of ± 2.066). This allows comparison between results with different error rates by correcting the error rate to 4% (MacKenzie 1992). Throughput based on effective measures is not only independent of speed-accuracy trade-offs (MacKenzie and Isokoski 2008) but also less dependent on device noise thereby enable more reliable comparisons (Teather and Stuerzlinger 2011; Li et al. 2018).

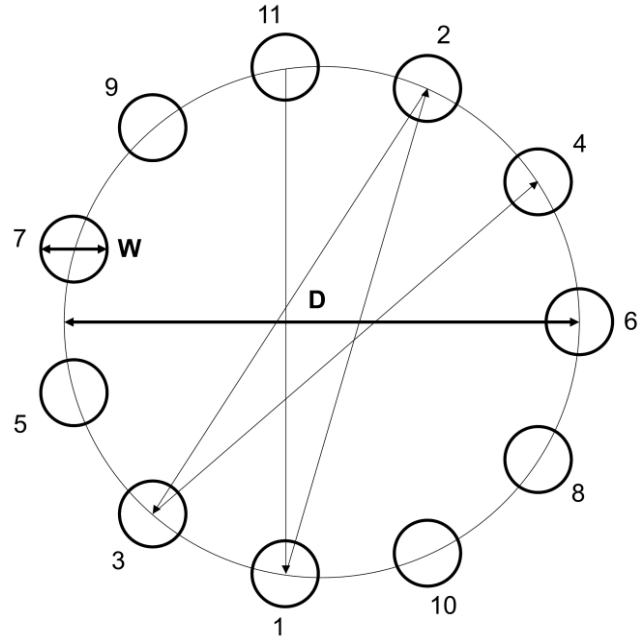

**Fig. 1.** Multi-directional tapping task (ISO 9241-9) with 11 targets. The arrows show the task axis participants follow to select targets alternating clockwise.

*2.2. ViewfinderVR*

Fig. 2 illustrates the operation procedure of our proposed technique, ViewfinderVR. ViewfinderVR aims to aid the selection of distant objects by utilizing a virtual panel that works somewhat similarly to the viewfinder of a digital camera, taking advantage of the *Through-the-lens* metaphor (Stoev and Schmalstieg 2002). The ViewfinderVR technique consists of three procedures: pre-configuration, configuration, and post-configuration. Before configuration, a virtual panel is located in front of users and follows their head movement, while their view is being reflected on the panel. After users place the distant targets inside the panel by moving and rotating their head, they can capture the panel image to fixate the view by reaching their non-dominant hand to the panel and grab it by pinching with the index finger. Once the panel is grabbed, it can be placed at any desired position and orientation. Furthermore, the panel view or the panel itself can be enlarged or shrunk by moving the hand up or down while pinching with the middle or ring finger, respectively. Users can always reset the panel setting and return it to the pre-configuration status by turning their non-dominant hand over and looking at the palm.

After configuration is completed, users can choose distant targets by selecting the targets reflected inside the panel view instead of selecting the actual targets. The panel cursor is coupled with the control cursor which moves on the control plane located in front of the view origin. The cursor for selection is rendered on the intersection point between the target object and the ray from the view origin to the control cursor. The panel cursor can be controlled via two methods: ray (ViewfinderVR-Ray) and touch (ViewfinderVR-Touch). ViewfinderVR-Ray works the same way as the traditional raycasting method (Facebook 2021) but the ray from the dominant hand interacts with the panel to determine the position of the panel cursor instead of the target object. Once the rendered cursor is located on the target object, users can select the object with a thumb-index pinch gesture. On the other hand, ViewfinderVR-Touch locates the panel cursor on the intersection point between the panel and the ray originating from the index fingertip of the dominant hand and directed perpendicular to the panel. The selection occurs when the index fingertip touches the panel surface.

In this study, the panel had a size of 0.4 × 0.3 × 0.01m and was located 0.4m away and 0.15m below from the head with an inclination of 30° before configuration. Note that the default panel setup was determined based on a series of internal pilot tests considering various circumstances such as field of view and hand tracking range of the VR headset, occlusion of sight caused by the panel, and location to comfortably reach the panel (Shin and Zhu 2011; Penumudi et al. 2020). The virtual hands were semi-transparent during the whole experiment to prevent occlusion of the object behind the hands (Fig. 2).

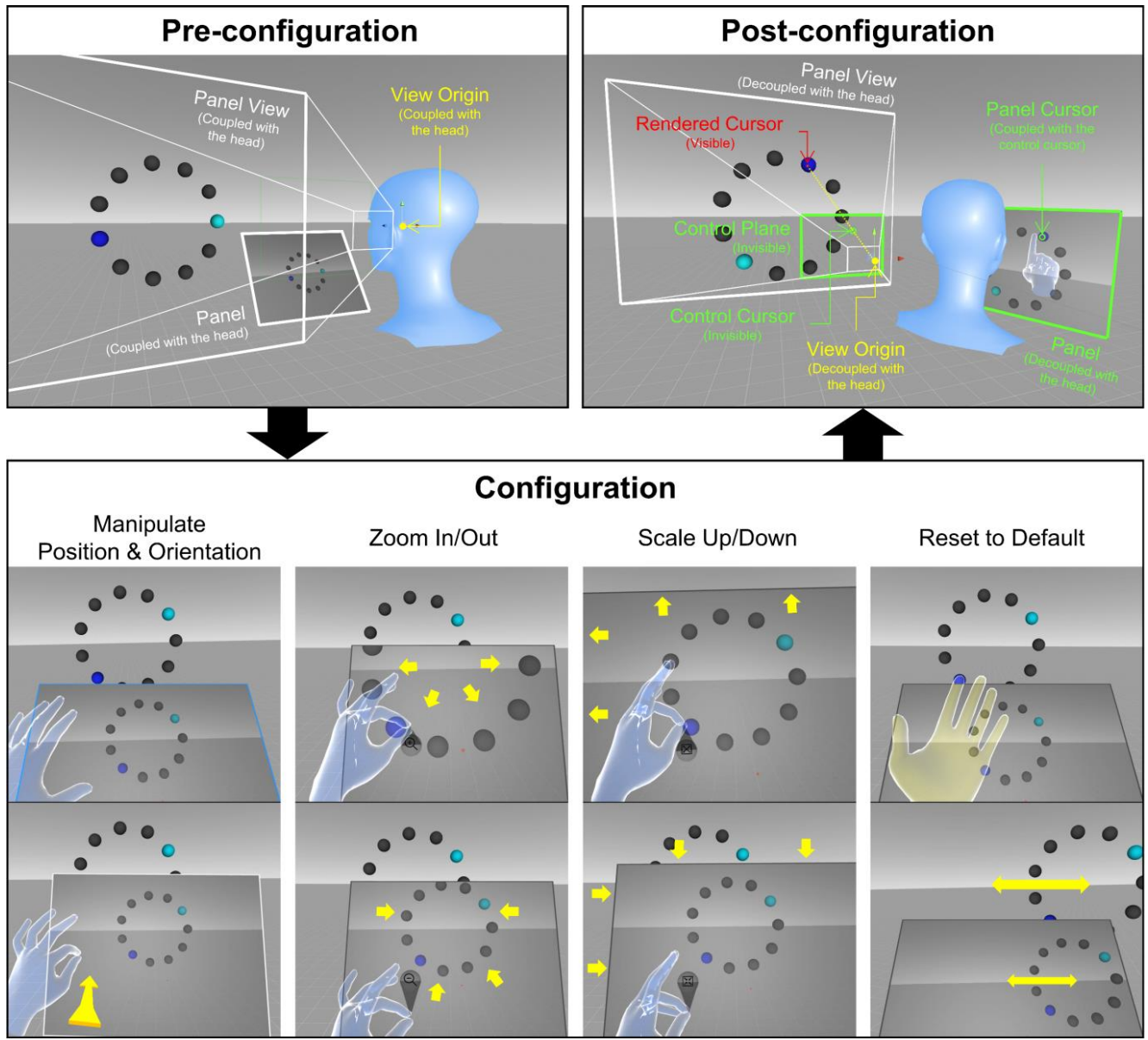

**Fig. 2.** The operation procedure of ViewfinderVR. A video demonstration of configuration options can be seen at the following link: https://vimeo.com/562391685.

## 3. Experiment

### 3.1. Participants

20 Korean young adults (12 males and 8 females) with a mean age of 24.0 (SD=4.1) and with normal or corrected to normal vision participated in the experiment. All participants were right-handed. 18 participants have experienced any kind of headset-based VR applications before this experiment, but 16 among them used the VR headset no more than once a year, showing the majority of participants were light VR users. 6 participants reported that they experienced mid-air interaction with the tracked virtual hand in VR. All participants gave consent for the experiment protocol approved by the University Institutional Review Board (IRB NO.: KH2021-009).

### 3.2. Experimental Settings

The participants were equipped with the Oculus Quest (resolution: 1440×1600 per eye; refresh rate: 72 Hz) VR headset. The headset was connected to the PC with an Intel Core i7-7700 processor running at 3.6 GHz, 16 GB of RAM, and an NVIDIA GeForce GTX 1080 Ti GPU, running Windows 10 through Oculus Link via a compatible USB 3.0 cable. Hands were tracked in real-time by the four fisheye monochrome cameras embedded in the Oculus Quest headset (Han et al. 2020), so the background was covered by a black screen fence to prevent any potential deterioration of tracking performance (Fig. 3). The distance between the participant and the screen fence was 1.2m, and no physical interruptions were caused by the screen fence during the whole experiment. The virtual environment used for the experiment was developed using Unity 2019.4.15f1.

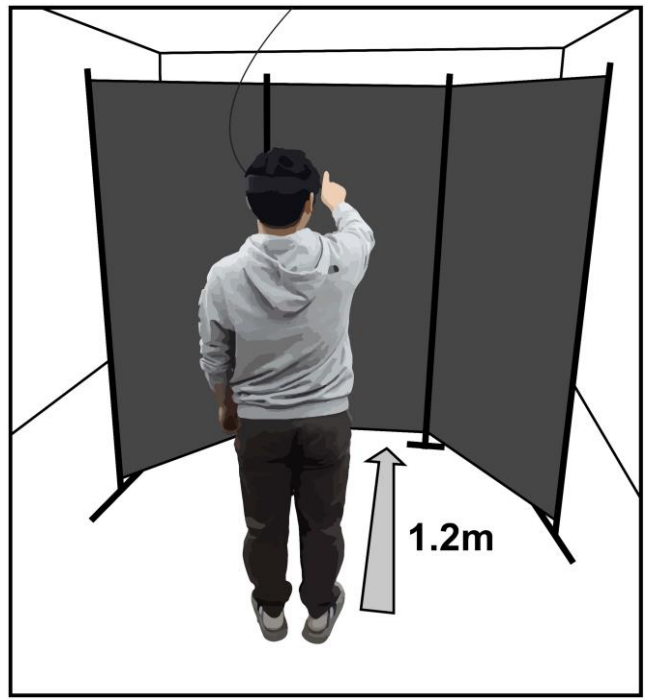

**Fig. 3.** Experimental settings in the real environment

### 3.3. Experimental Design and Procedure

First, participants filled in a pre-test questionnaire asking about their demographic information and prior experience with VR. Then, they put on the VR headset and performed a multidirectional tapping task described in the ISO 9241-9 standard (ISO 2002; Soukoreff and MacKenzie 2004). In this task, 11 dark gray circular targets were placed in a circular pattern, and participants were asked to select the target highlighted in blue as accurately and quickly as possible while prioritizing accuracy over speed. The subsequent target and the missed target were highlighted in cyan and magenta, respectively. A sequence of trials (one circle) consisted of 11 target selections, and total 3 sequences were conducted for each of 3 techniques: Raycasting, Viewfinder-Ray, and Viewfinder-Touch (Fig. 4-a) × 2 target sizes: large and small (0.2619m and 0.0873m; represented as the visual angle of 3° and 1°, respectively) × 2 target distances: short and long (1m and 2m) within-subject conditions. A combination of 2 target sizes and distances resulted in 4 target conditions with different *ID*s: 2.27, 3.11, 3.64, and 4.58 (Fig. 4-b), and target depth was kept 5m in all sessions. In each test condition, participants were allowed to practice at first as much as they want until they fully understand and feel confident about the technique, followed by 30 seconds of rest before the test session. For ViewfinderVR techniques, participants were asked to freely explore and determine the panel setup that can achieve the best task performance during the practice session, and then reconfigure the panel from the default status to measure the time needed for configuration. During the test session, participants were not allowed to reconfigure the panel. After all trials in each test condition were finished, participants gave ratings to NASA-TLX (Hart and Staveland 1988). Whenever the selection occurred, the participants were informed with a click sound.

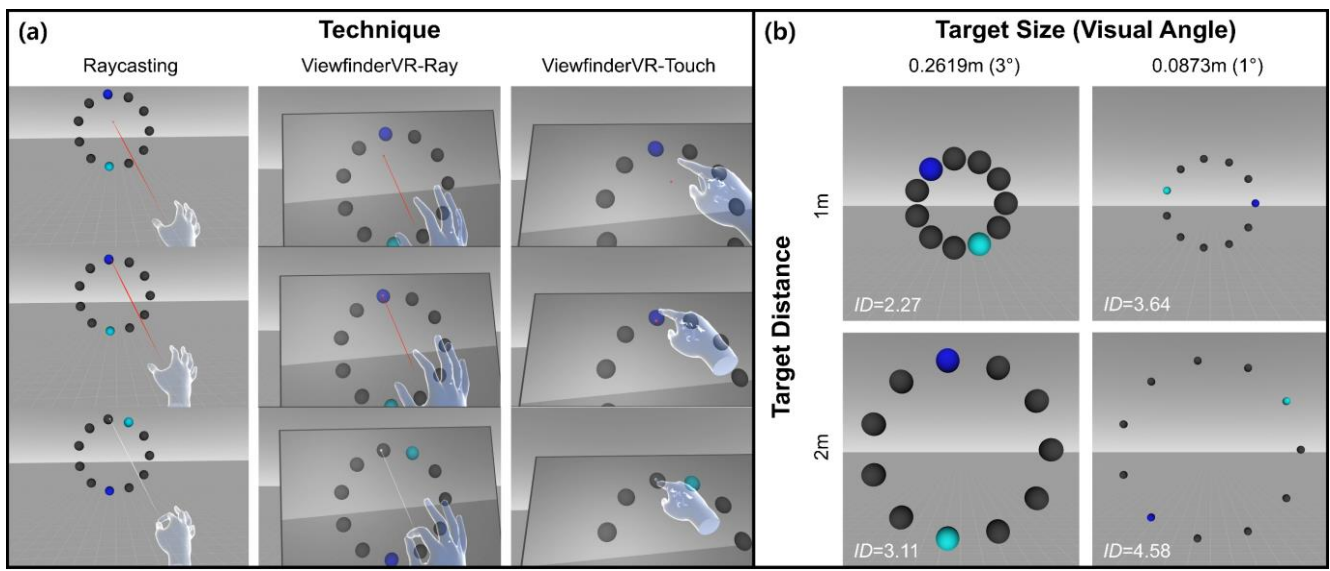

**Fig. 4.** (a) Three techniques and (b) four target conditions used in this study. A video demonstration of techniques can be seen at the following link: https://vimeo.com/560739540.

### 3.4. Data Analysis

Out of 7920 trials (20 participants × 12 test conditions × 3 sequences × 11 trials), 89 trials (1.1%) with the distance between the cursor and the target center at selection was larger than 3 times of target radius were classified as outliers and were excluded from the analysis. These outliers were caused by consecutive pinch detection due to participants' occasional mistakes or hand tracking errors. After the outlier removal, 3 performance measures: movement time, error rate, and throughput (MacKenzie 2015) and 7 NASA-TLX ratings for perceived workload: mental demand, physical demand, temporal demand, performance, effort, frustration, and weighted rating were collected and arranged for the analysis of variance (ANOVA) where the variation from participants was blocked. In addition, 4 user behavior measures: hand movement, head movement, actual target depth, and adjusted target visual angle were calculated and analyzed to have a better understanding of the results. Hand and head movements were defined as the length of movement trajectory of the tracked dominant hand and the VR headset, respectively. Actual target depth was defined as the measured distance between the VR headset and center of the target group, whereas adjusted target visual angle was defined as the visual angle of the interacted target calculated based on the actual target depth. For ViewfinderVR techniques, actual target depth indicated the distance between the VR headset and center of the panel, and adjusted visual angle was calculated for the target proxies reflected inside the virtual panel. Matlab R2019a and Minitab 19 were used to conduct all data processing and statistical analyses at a significance level of 0.05, and the effect size in terms of eta-squared ($\eta^2 = SS_{Treatment}/SS_{Total}$) was further calculated to check practical significance. A general rule on the magnitudes of the effect size with $\eta^2$ is as follows: small-$\eta^2 \sim 0.01$, medium-$\eta^2 \sim 0.06$ and large-$\eta^2 \sim 0.14$ (Cohen 1988).

## 4. Results

We only highlight a subset of significant results related to the technique for better clarity. The entire descriptive statistics and ANOVA results are presented in the Appendix section.

### 4.1. Performance

A significant main effect of the technique was observed on all performance measures (Fig. 5): movement time ($F_{2,689} = 317.38$, $p < 0.001$, $\eta^2 = 0.243$), error rate ($F_{2,689} = 34.20$, $p < 0.001$, $\eta^2 = 0.048$), and throughput ($F_{2,689} = 451.07$, $p < 0.001$, $\eta^2 = 0.408$). The movement time was the shortest at ViewfinderVR-Touch ($M = 1.29$s, $SD = 0.10$), followed by ViewfinderVR-Ray ($M = 1.69$s, $SD = 0.60$), and then Raycasting ($M = 2.38$s, $SD = 1.13$). Likewise, ViewfinderVR-Touch had the highest throughput ($M = 3.16$ bit/s, $SD = 0.97$), followed by ViewfinderVR-Ray ($M = 2.31$ bit/s, $SD = 0.62$), and then Raycasting ($M = 1.64$ bit/s, $SD = 0.61$). For error rate, Raycasting ($M = 17.08\%$, $SD = 19.28$) was significantly higher than ViewfinderVR-Ray ($M = 9.3\%$, $SD = 12.42$) and ViewfinderVR-Touch ($M = 10.36\%$, $SD = 13.79$), where no significant difference between two ViewfinderVR techniques was found according to the post hoc test.

Fig. 6 shows performance measures by technique, target size, and target distance. A significant interaction effect of technique × target size ($F_{2,689} = 43.12$, $p < 0.001$, $\eta^2 = 0.033$) was found for movement time. Movement time with small targets increased compared to large targets in all techniques, but Raycasting showed a larger increase compared to the ViewfinderVR techniques. For error rate, significant interaction effects of technique × target size ($F_{2,689} = 20.22$, $p < 0.001$, $\eta^2 = 0.028$), technique × distance ($F_{2,689} = 3.37$, $p = 0.035$, $\eta^2 = 0.005$), and technique × target size × target distance ($F_{2,689} = 4.19$, $p = 0.016$, $\eta^2 = 0.006$) were found. Error rate of Raycasting was significantly larger than ViewfinderVR-Touch but not ViewfinderVR-Ray at large targets. When it came to small targets, Raycasting had a significantly larger error rate than both ViewfinderVR techniques when target distance was short, but larger than ViewfinderVR-Ray only when target distance was long due to the sudden increase of error in ViewfinderVR-Touch. For throughput, interaction effects of technique × target size ($F_{2,689} = 5.59$, $p = 0.004$, $\eta^2 = 0.005$) and technique × target distance ($F_{2,689} = 3.50$, $p = 0.031$, $\eta^2 = 0.003$) were found to be significant. As with the movement time, a steeper decrease of throughput from large to small targets at Raycasting in comparison with the ViewfinderVR techniques was observed. In addition, the throughput of the ViewfinderVR techniques decreased when target distance became longer at small targets, while throughput remained similar at Raycasting. Fig. 7 illustrates cursor movement trajectories by technique, target size, and target distance.

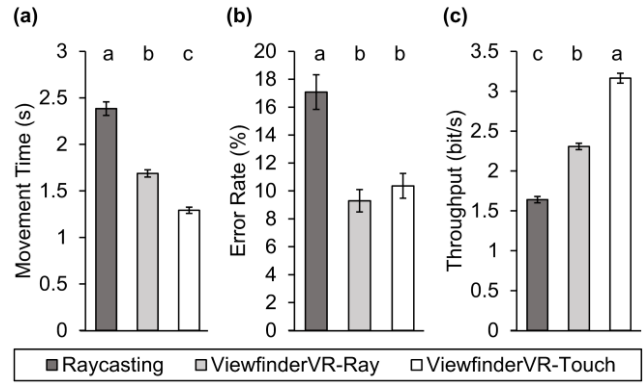

**Fig. 5.** Mean (+SE) (a) movement time, (b) error rate, and (c) throughput by technique. *Note:* Alphabetical letters above bars represent post hoc grouping.

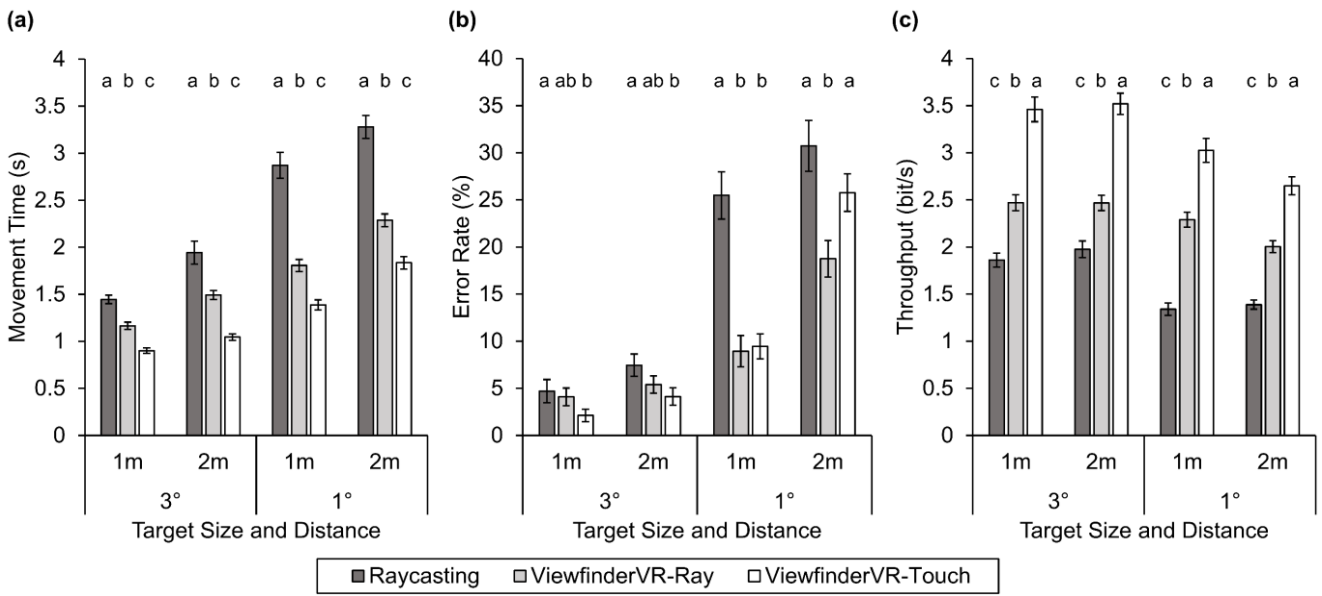

**Fig. 6.** Mean (+SE) (a) movement time, (b) error rate, and (c) throughput by technique, target size, and target distance. *Note:* Alphabetical letters above bars represent post hoc grouping for techniques in each target size and distance.

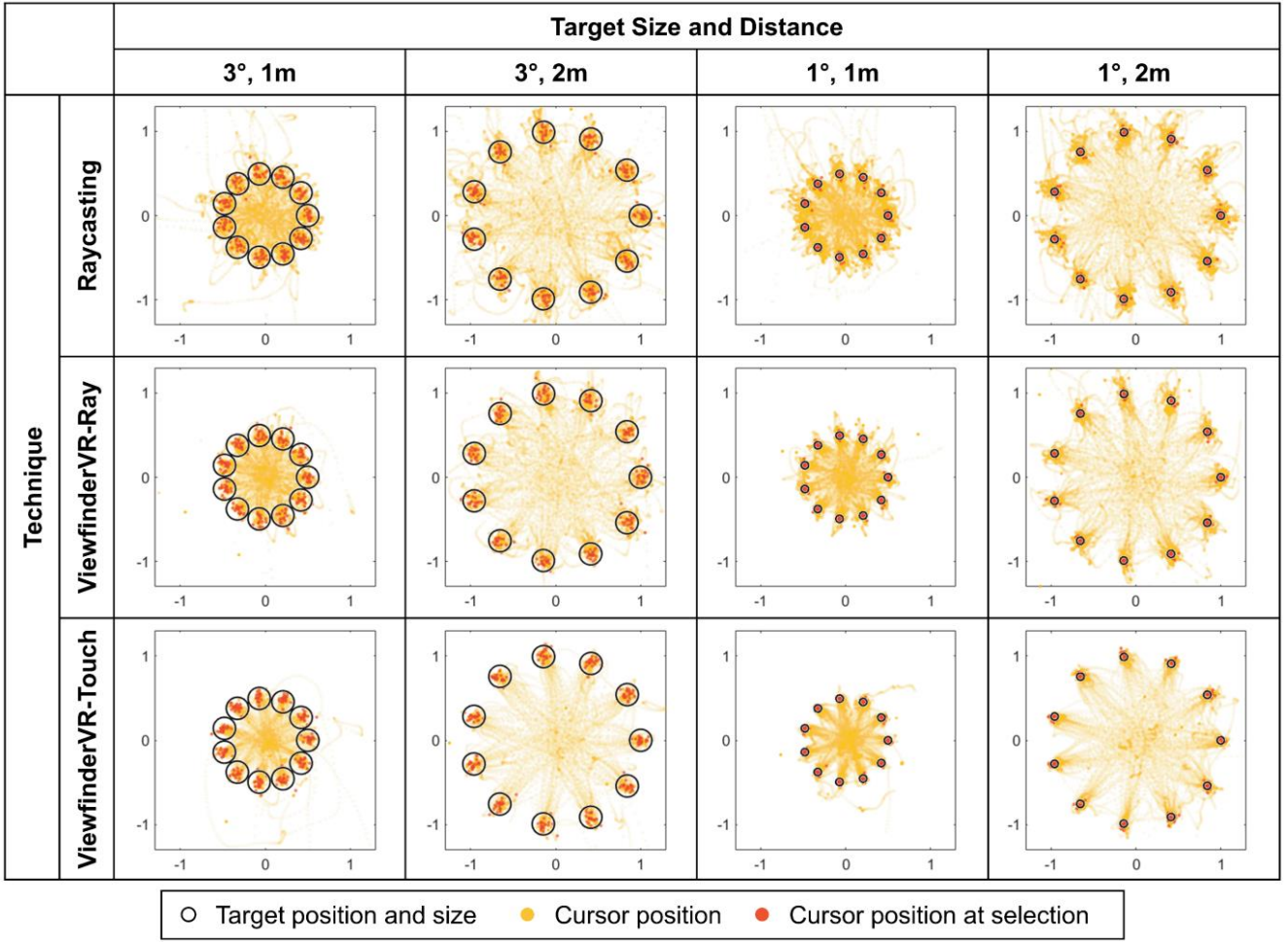

**Fig. 7.** Scatter plots of the cursor position by technique, target size, and target distance. *Note:* The unit of axes is in meter (Unity scale).

### 4.2. Perceived Workload

A significant main effect of the technique was observed on all perceived workload measures (Fig. 8): mental demand ($F_{2,209} = 60.74$, $p < 0.001$, $\eta^2 = 0.153$), physical demand ($F_{2,209} = 39.96$, $p < 0.001$, $\eta^2 = 0.112$), temporal demand ($F_{2,209} = 39.76$, $p < 0.001$, $\eta^2 = 0.105$), performance ($F_{2,209} = 21.43$, $p < 0.001$, $\eta^2 = 0.072$), effort ($F_{2,209} = 52.15$, $p < 0.001$, $\eta^2 = 0.127$), frustration ($F_{2,209} = 30.29$, $p < 0.001$, $\eta^2 = 0.084$), and weighted rating ($F_{2,209} = 55.38$, $p < 0.001$, $\eta^2 = 0.127$). ViewfinderVR-Touch had the lowest weighted NASA-TLX rating ($M = 33.71$, $SD = 25.26$), followed by ViewfinderVR-Ray ($M = 43.22$, $SD = 25.07$), and then Raycasting ($M = 57.05$, $SD = 25.52$). All raw NASA-TLX workload ratings had similar results where Viewfinder-Touch was the lowest, Raycasting was the highest, and Viewfinder-Ray stayed in between, although no significant difference between the two ViewfinderVR techniques was found in frustration.

Fig. 9 shows NASA-TLX workload ratings by technique, target size, and target distance. For raw ratings, a significant effect of technique × target size was found at all subscales except effort: mental demand ($F_{2,209} = 4.82$, $p = 0.009$, $\eta^2 = 0.012$), physical demand ($F_{2,209} = 3.49$, $p = 0.032$, $\eta^2 = 0.010$), temporal demand ($F_{2,209} = 3.95$, $p = 0.021$, $\eta^2 = 0.010$), performance ($F_{2,209} = 8.56$, $p < 0.001$, $\eta^2 = 0.029$), and frustration ($F_{2,209} = 4.90$, $p = 0.008$, $\eta^2 = 0.014$). For weighted rating, significant interaction effects of technique × target size ($F_{2,209} = 5.91$, $p = 0.003$, $\eta^2 = 0.014$) and technique × target size × target distance ($F_{2,209} = 3.63$, $p = 0.028$, $\eta^2 = 0.008$) were found. A similar tendency was found across all workload ratings that the rating of Raycasting was generally higher yet occasionally similar compared to the ViewfinderVR techniques at large targets but the discrepancy became more obvious at small targets.

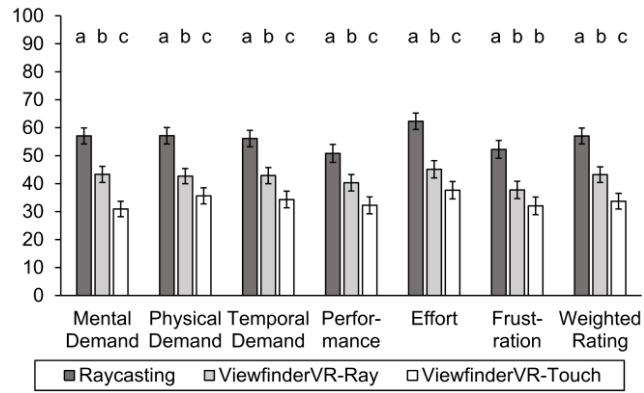

**Fig. 8.** Mean (+SE) NASA-TLX ratings by technique. *Note:* Alphabetical letters above bars represent post hoc grouping.

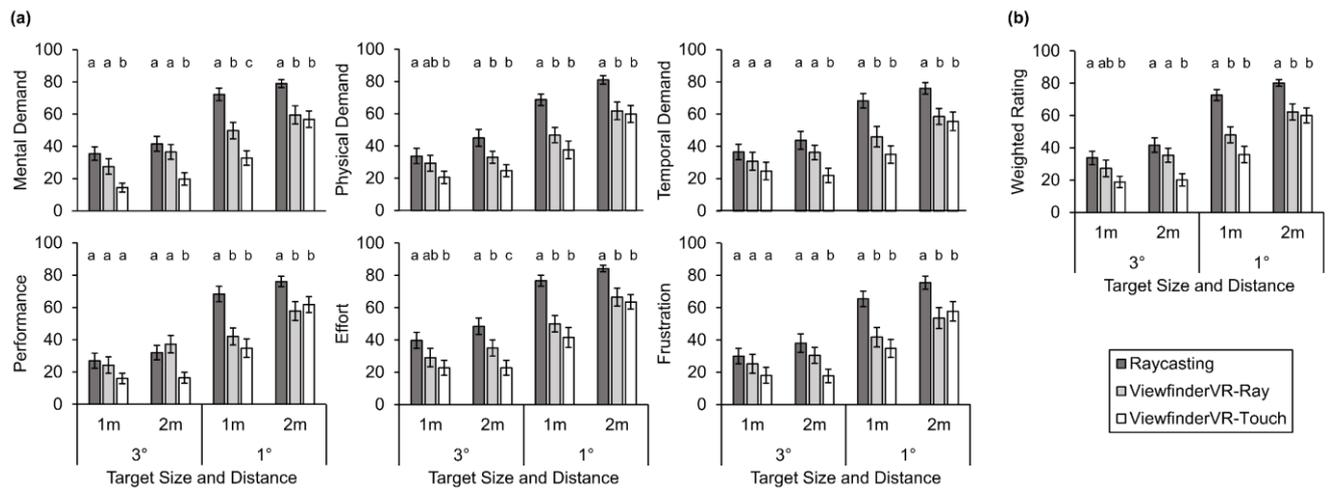

**Fig. 9.** Mean (+SE) (a) raw and (b) weighted NASA-TLX ratings by technique, target size, and target distance. *Note:* Alphabetical letters above bars represent post hoc grouping for techniques in each target size and distance.

### 4.3. User Behavior

There was a significant main effect of the technique on all user behavior measures (Fig. 10): hand movement ($F_{2,689} = 112.66$, $p < 0.001$, $\eta^2 = 0.207$), head movement ($F_{2,689} = 79.87$, $p < 0.001$, $\eta^2 = 0.059$), actual target depth ($F_{2,689} = 334002.29$, $p < 0.001$, $\eta^2 = 0.998$), and adjusted target visual angle ($F_{2,689} = 290.75$, $p < 0.001$, $\eta^2 = 0.171$). The hand and head movements of ViewfinderVR-Touch (Hand: $M = 3.64$m, $SD = 2.33$; Head: $M = 0.50$m, $SD = 0.32$) was significantly longer than ViewfinderVR-Ray (Hand: $M = 1.77$m, $SD = 1.62$; Head: $M = 0.34$m, $SD = 0.24$) and Raycasting (Hand: $M = 1.61$m, $SD = 1.34$; Head: $M = 0.36$m, $SD = 0.24$). Understandably, the actual target depth of Raycasting remained very close to the depth of original targets ($M = 5.00$m, $SD = 0.09$), exceeding the ones of ViewfinderVR techniques by a large margin. The panel of ViewfinderVR-Touch ($M = 0.40$m, $SD = 0.06$) was placed slightly closer to the user compared to ViewfinderVR-Ray ($M = 0.47$m, $SD = 0.11$). For adjusted target visual angle, Viewfinder-Touch was the largest ($M = 3.68°$, $SD = 1.80$), followed by ViewfinderVR-Ray ($M = 3.37°$, $SD = 1.91$), and then Raycasting ($M = 2.00°$, $SD = 1.00$).

Fig. 11 shows user behavior measures by technique, target size, and target distance. A significant interaction effect of technique × target distance ($F_{2,689} = 5.18$, $p = 0.006$, $\eta^2 = 0.010$) was found for hand movement, whereas significant interaction effects of technique × target size ($F_{2,689} = 22.49$, $p < 0.001$, $\eta^2 = 0.017$), technique × target distance ($F_{2,689} = 23.13$, $p = 0.035$, $\eta^2 = 0.017$), and technique × target size × target distance ($F_{2,689} = 9.70$, $p < 0.001$, $\eta^2 = 0.007$) were found for head movement. In both movement measures, a significant increment at ViewfinderVR-Touch for targets with small size and long distance was identified. While there was no significant interaction effect for actual target depth, interaction effects of technique × target size ($F_{2,689} = 11.16$, $p < 0.001$, $\eta^2 = 0.007$), technique × target distance ($F_{2,689} = 46.14$, $p < 0.001$, $\eta^2 = 0.027$), and technique × target size × target distance ($F_{2,689} = 4.33$, $p = 0.014$, $\eta^2 = 0.003$) were significant on adjusted target visual angle. Expectably, the target visual angle of Raycasting could not be adjusted thus stayed almost identical to the visual angle of original targets. On the contrary, the target visual angle of ViewfinderVR techniques was adjustable thereby configured to be significantly larger than Raycasting for improved selection performance. Fig. 12 depicts hand, head movement trajectories, and panel placements by technique, target size, and target distance.

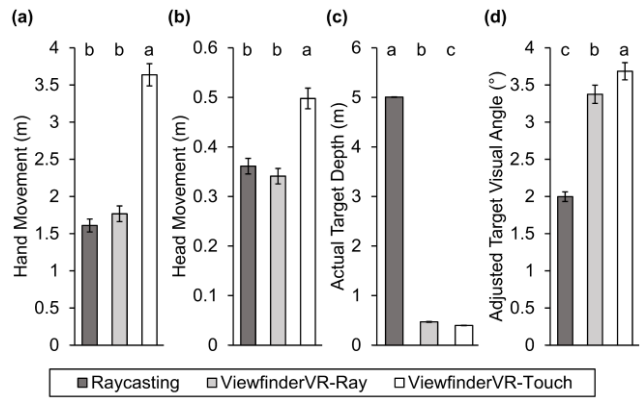

**Fig. 10.** Mean (+SE) (a) hand movement, (b) head movement, (c) actual target depth, and (d) adjusted target visual angle by technique, target size, and target distance. *Note:* Alphabetical letters above bars represent post hoc grouping.

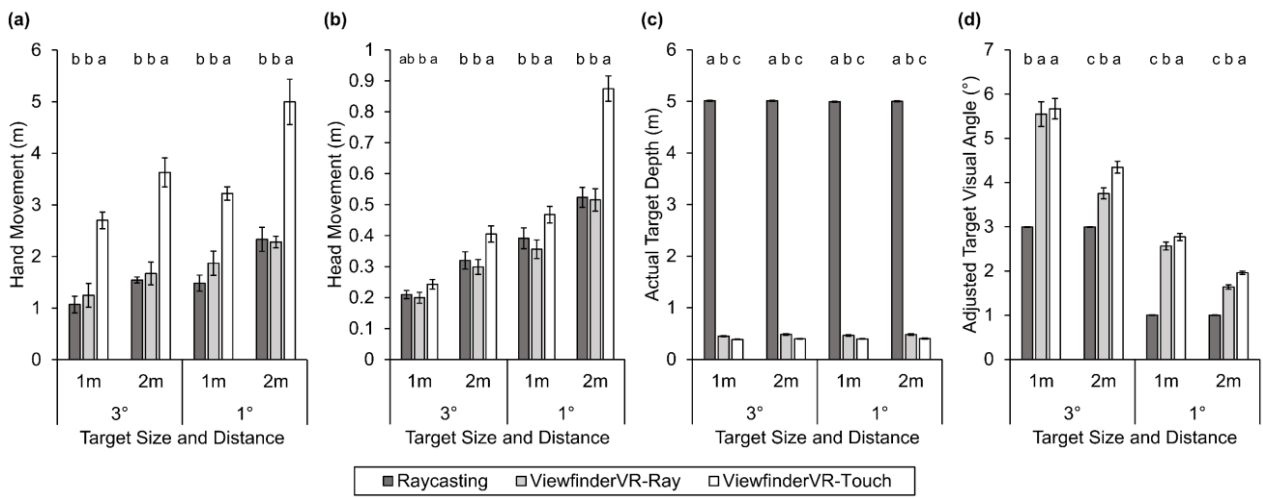

**Fig. 11.** Mean (+SE) (a) hand movement, (b) head movement, (c) actual target depth, and (d) adjusted target visual angle by technique, target size, and target distance. *Note:* Alphabetical letters above bars represent post hoc grouping for techniques in each target size and distance.

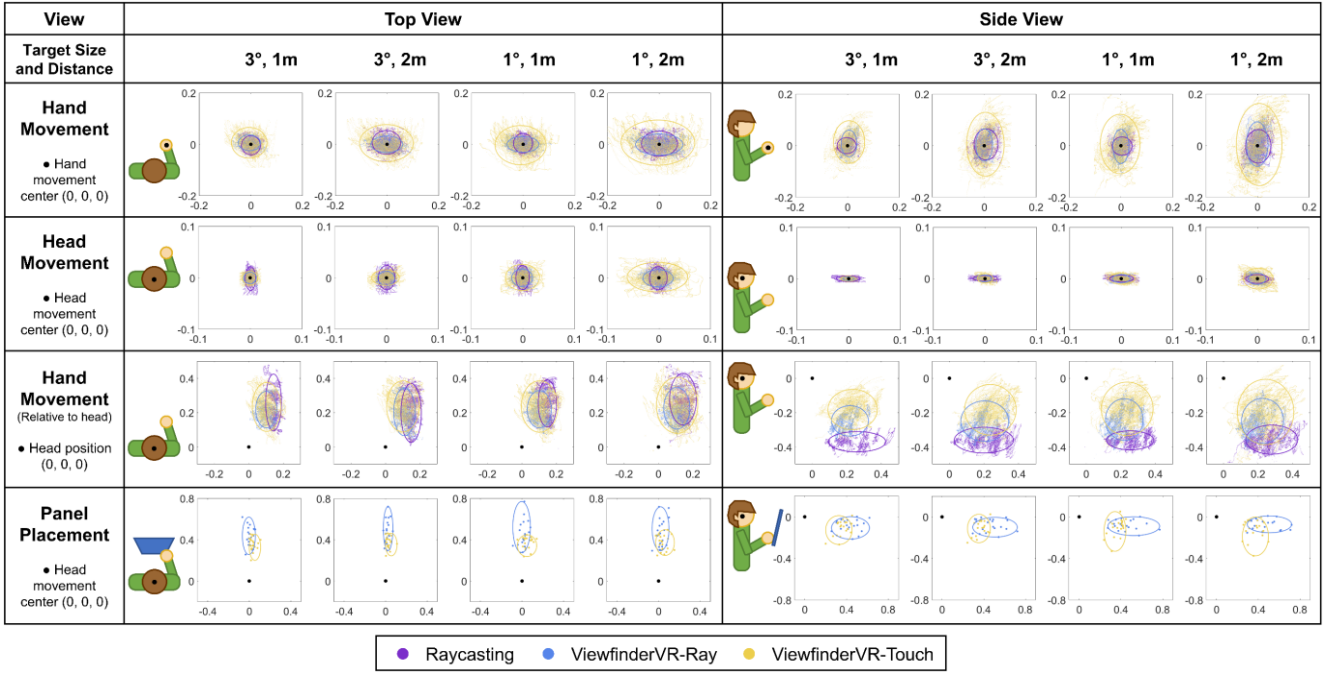

**Fig. 12.** Scatter plots and 95% percentile ellipses of hand, head, and panel positions by technique, target size, and target distance. *Note:* Hand and head movement centers are defined as the average point of all hand and head positions in a single sequence of trials, respectively. The unit of axes is in meter (Unity scale).

## 5. Discussion

### 5.1. Performance

The result showed that our proposed technique, ViewfinderVR could significantly outperform Raycasting in every performance aspect including movement time, error rate, and throughput. To be more specific, ViewfinderVR-Ray lowered movement time by 20.0–36.9%, error rate by 13.0–64.9%, and elevated throughput by 24.7–70.9%, while ViewfinderVR-Touch lowered movement time by 37.9–51.6%, error rate by 16.1–63.0%, and elevated throughput by 77.8–126.1% compared to Raycasting depends on the target condition. The superior performance of the ViewfinderVR techniques appears to have a strong connection with the actual target depth and adjusted target visual angle (Fig. 11). Numerous studies have demonstrated the effects of the display size and distance to the user, which determine the target visual angle, on task performance. For mouse interaction, Hourcade and Bullock-Rest (2012) reported a negative impact of large target depth on performance with a steep decline for targets with visual angles below 0.05°, whereas Browning and Teather (2014) found that the small display size reduced throughput by around 20%. For touch interaction, the small display size induced a high error rate and was perceived as imprecise and difficult to use (List and Kipp 2019). By taking advantage of the ViewfinderVR technique, the target visual angle of ViewfinderVR-Ray and ViewfinderVR-Touch could be adjusted to be 25.4–156.0% and 45.2–177.0% larger than the one of Raycasting, respectively.

The approach to benefit selection by enlarging the targets is not completely new. The target expansion technique, which allowed targets to expand dynamically as the cursor approaches (McGuffin and Balakrishnan 2005), was proven to enhance performance using a mouse (Hwang et al. 2013) and a touch display (Yang et al. 2011; Perea et al. 2020). With a VR headset, Yang et al. (2019) applied a similar strategy to magnify keys in a virtual keyboard and found enhancement in the performance of VR text entry. The target expansion technique is designed to be mainly applied in user interfaces where targets, target groups, or points of interest are pre-defined or need to be configured. On the other hand, ViewfinderVR focuses on the application in a 3D environment where the position, orientation, and scale of any space within a view can be manipulated. It is worth noting that ViewfinderVR is also capable of downscaling the view which could enable less fatiguing and time-consuming selection by reducing the distance between targets.

According to the result, ViewfinderVR-Touch outperformed ViewfinderVR-Ray in terms of movement time (a decrease of 20.1–30.2%) and throughput (an increase of 32.3–42.5%). Direct touch interaction was known to provide easy access to contents with an intuitive way of interacting with objects. Studies have shown users performed the task faster and more accurately with higher satisfaction with direct touch input than with indirect mouse input (Sears and Shneiderman 1991; Kin et al. 2009; Watson et al. 2013). Our result further supports that more efficient control of the cursor was available with the direct approach (ViewfinderVR-Touch) compared to the indirect ray-based approaches (Fig. 7). The difference in triggering the selection also might take part in improving the performance of ViewfinderVR-Touch. ViewfinderVR-Touch requires users to thrust their index finger towards the panel and the selection is triggered when the fingertip penetrates the panel surface, while ray-based techniques require users to perform an additional gesture of pinch for confirmation of the selection. This pinch gesture will disturb the tracked hand position and orientation thus induce a slight jitter of the cursor at the moment of selection, commonly referred to as the Heisenberg effect of spatial interaction (Bowman et al. 2001). This is similar to the concern introduced in the work of Zhou et al. (2020) that selection was confirmed when the finger and stimulus intersected to avoid a need for an additional operation that causes the Heisenberg effect. According to the study of Lin et al. (2019), the index thrust gesture turned out to have a higher preference, faster selection speed, and fewer errors for selecting 2D targets compared to the palm click gesture which shares similarities with the pinch gesture we used in this study.

One noteworthy result is that the error rate of ViewfinderVR-Touch was higher than ViewfinderVR-Ray for targets with small size and long distance. This is because enlarging the virtual panel or its view results in increasing not only target size but also target distance, indicating "the larger the better" does not apply here as a longer target distance leads to longer movement time (Wang et al. 2013). Hence, the optimal balance between target size and distance should be found (Lee et al. 2013). When the target size was small, participants had to increase the size of the panel or its view to be large enough to facilitate a more accurate selection. As ViewfinderVR-Touch required users to reach their hand to each of distanced targets, considerably larger movements in the upper body were caused especially when the distance between targets was long (Fig. 11 and Fig. 12). Larger body movements appear to pose more difficulties in gesture control and increase the occurrence of errors (Tao et al. 2021). This result was consistent with Lou et al. (2018), in which a longer target distance increased the physical movements of users and damaged the selection efficiency of freehand remote pointing on a large display.

## 5.2. Perceived Workload

Mid-air interactions are prone to fatigue as the user's arm is suspended in mid-air while interacting with objects, also known as the gorilla arm effect (Boring et al. 2009); therefore, physical demand becomes a critical issue when it comes to long-term interaction. The significantly lower NASA-TLX workload ratings of the ViewfinderVR techniques in comparison with Raycasting denotes workload could be relieved by enlarging the target visual angle. This is consistent with the previous studies that reported higher perceived workload and biomechanical stress in the neck and shoulders when interacting with smaller targets in VR and AR (Kim et al. 2020; Kia et al. 2020, 2021). The effect of the target location should also be noted. Penumudi et al. (2020) warned excessive vertical target locations (15° above and 30° below eye height) could induce musculoskeletal discomfort during VR interactions, while Kim et al. (2020) recommended the target depth of 0.3–0.6 m to reduce the biomechanical load for AR interactions. In the current study, participants placed the virtual panel to be 0.47m away and 10.0° below them for ViewfinderVR-Ray and 0.40m away and 17.2° below them for ViewfinderVR-Touch on average.

Interestingly, participants felt ViewfinderVR-Touch was less physically demanding than the ray-based techniques, although ViewfinderVR-Touch induced significantly larger hand and head movements and higher hand placement (Fig. 11 and Fig. 12). It can be assumed this result was mainly caused by the difference in task completion time. Fatigue in mid-air interaction is heavily influenced by the time for interaction that the upper arm muscles remain active, as reflected in biomechanical models to quantify arm fatigue of mid-air interactions such as consumed endurance (Hincapié-Ramos et al. 2014) and cumulative fatigue model (Jang et al. 2017; Liu et al. 2018). The movement time of ViewfinderVR-Touch was 37.9–51.6% shorter than Raycasting and 20.1–30.2% shorter than ViewfinderVR-Ray depends on the target condition, showing that participants could hold up their arms for a shorter period for performing the same amount of tasks when using ViewfinderVR-Touch. Furthermore, some evidence has indicated larger body movement does not always result in a larger physical burden. Nunnari et al. (2016) verified that postural variability could distribute muscular loads to different body parts during mid-air gestural interaction, demonstrating a 19% decrease in simulated average musculoskeletal loads on the shoulder, neck, and back area in their user study. Accordingly, the rating of physical demand of ViewfinderVR-Touch was 26.2–45.5% lower than Raycasting and 3.2–29.9% lower than ViewfinderVR-Ray in this study.

Participants gave lower workload ratings for ViewfinderVR-Touch on not only physical demand but also other NASA-TLX subscales. This result is somewhat expected since studies have proved the better performance and less required amount of mental resources on using direct input techniques compared to indirect counterparts (Charness et al. 2004; Murata and Iwase 2005; McLaughlin et al. 2009; Brasier et al. 2020). Indirect input techniques require cognitive processing of spatial translation between the physical movement of the human body and the virtual movement of the cursor which affects attentional requirements (Wickens 1980). In contrast, direct input techniques have no intermediary and require users to directly touch objects by bringing their hands closer as in ViewfinderVR-Touch. Hence, direct input techniques provide some relative benefits to users with deteriorated attention skills such as the elderly (Charness et al. 2004; McLaughlin et al. 2009; Jochems et al. 2013) and patients with dementia (Andrade Ferreira et al. 2020), thus such users may benefit more with ViewfinderVR-Touch in the same manner. Although direct touch interaction forces users to move their hands more, it can further alleviate cognitive demand, as keeping hands proximate to objects is advantageous for processing and learning visuospatial information (Reed et al. 2006; Tseng and Bridgeman 2011; Brucker et al. 2021). Moreover, lesser fatigue might also play a role in lowering workload ratings of mental demand and other associated subscales, as shoulder muscular fatigue from static posture is known to concurrently reduce cognitive attentional resources (Stephenson et al. 2020).

### 5.3. *Limitations and Future Work*

Some limitations of the proposed technique should be noted. Firstly, ViewfinderVR requires extra time and effort to configure the panel to improve the object selection. In this study, participants required 21.1 seconds on average (SD = 11.8) for panel configuration. For this reason, ViewfinderVR is not appropriate for scenarios where the view needs to be changed frequently, since the user has to reconfigure the panel whenever the target beyond the panel view needs to be selected. Although the selection can occur at the pre-configuration stage, the performance deteriorates due to the repetitive shifts in the panel position and view caused by inherent head oscillations. Next, ViewfinderVR is not capable of enhancing the selection of occluded objects. Projecting the 3D environment onto a 2D panel avoids binocular parallax but omits the depth information, thus occluded objects cannot be selected similar to the case of traditional raycasting. Other techniques which specifically aimed for improvements in the selection of occluded objects or objects in a dense environment such as *DepthRay* (Grossman and Balakrishnan 2006), *Guidance Rays* (Xu et al. 2013), *RayCursor* (Baloup et al. 2019), and *vMirror* (Li et al. 2021) can be used in combination with the proposed technique to aid selection of such conditions, although effectiveness and potential conflicts need to be further validated.

There are several potential directions to further improve the proposed technique. First, the bimanual operation remains unexplored. As studies have shown speed advantages of bimanual touch over a pair of mice (Forlines et al. 2007; Kin et al. 2009), it is expected that bimanual interaction may benefit ViewfinderVR-Touch more compared to Raycasting. Second, multi-panel interaction is worthy to investigate. Some scenarios could benefit by placing multiple panels; for instance, multiple pre-configured panels widen the coverage of views and enable users to select objects in multiple spaces without the necessity to reconfigure the panel every time for each space. Third, the procedure of panel configuration can be minimized via simplification or automatization. The current technique requires the user to manually configure the panel by manipulating the position, orientation, and scale of the panel, in addition to the magnification of the panel view. One approach to tackle this issue could be to interact with the head-coupled panel large enough to cover the whole view, thereby make manipulation of the panel unnecessary. The other approach is to implement a strategy to automatically determine the panel configuration based on the user and environment characteristics such as field of view, target size and distance, and ergonomic cost (Evangelista Belo et al. 2021). Future work should explore ways to apply the aforementioned features to ViewfinderVR.

## 6. Conclusion

We have proposed a new technique named ViewfinderVR for the improved selection of distant objects in VR, which allows users to configure the interaction space projected onto a virtual panel in reach and to select objects inside the panel through ray-based or touch interaction. A user study was conducted to evaluate and compare the proposed technique with traditional raycasting through a 2D Fitts' law-based test. Experimental results showed that ViewfinderVR induced shorter movement time, lower error rate, higher throughput, and lower perceived workload compared to traditional raycasting for not only small but also large targets by reducing the target depth and enlarging the target visual angle. Between two selection methods in ViewfinderVR, touch interaction outperformed ray-based interaction in terms of movement time, throughput, and perceived workload with more efficient and accurate cursor control from the directness of interaction. However, touch interaction had a higher error rate for targets with small size and long distance, where excessive hand and head movements caused by the expanded target distance led to the increase of difficulties in gesture control hence error occurrence. The proposed technique is expected to benefit VR users to better select distant objects inside the virtual environment, even though caution is needed when the user should change the view frequently as reconfiguration of the panel is required each time the view is changed.

**Acknowledgements**

This work was supported by the Basic Science Research Program through the National Research Foundation of Korea funded by the Ministry of Science, ICT and Future Planning (NRF-2020R1F1A1048510) and the KAIST Faculty Research Fund (A0601003029).

# References

Andrade Ferreira LD, Ferreira H, Cavaco S, et al (2020) User Experience of Interactive Technologies for People With Dementia: Comparative Observational Study. JMIR Serious Games 8:e17565. https://doi.org/10.2196/17565

Andujar C, Argelaguet F (2007) Anisomorphic ray-casting manipulation for interacting with 2D GUIs. Comput Graph 31:15–25. https://doi.org/10.1016/j.cag.2006.09.003

Argelaguet F, Andujar C (2013) A survey of 3D object selection techniques for virtual environments. Comput Graph 37:121–136. https://doi.org/10.1016/j.cag.2012.12.003

Argelaguet F, Andujar C (2009) Visual feedback techniques for virtual pointing on stereoscopic displays. In: Proceedings of the 16th ACM Symposium on Virtual Reality Software and Technology - VRST '09. ACM Press, New York, New York, USA, p 163

Bacim F, Kopper R, Bowman DA (2013) Design and evaluation of 3D selection techniques based on progressive refinement. Int J Hum Comput Stud 71:785–802. https://doi.org/10.1016/j.ijhcs.2013.03.003

Baloup M, Pietrzak T, Casiez G (2019) Raycursor: A 3D Pointing Facilitation Technique based on Raycasting. In: Conference on Human Factors in Computing Systems - Proceedings

Batmaz AU, Seraji MR, Kneifel J, Stuerzlinger W (2021) No Jitter Please: Effects of Rotational and Positional Jitter on 3D Mid-Air Interaction. In: Arai K, Kapoor S, Bhatia R (eds) Proceedings of the Future Technologies Conference (FTC) 2020, Volume 2. FTC 2020. Advances in Intelligent Systems and Computing, vol 1289. pp 792–808

Boring S, Jurmu M, Butz A (2009) Scroll, tilt or move it. In: Proceedings of the 21st Annual Conference of the Australian Computer-Human Interaction Special Interest Group on Design: Open 24/7 - OZCHI '09. ACM Press, New York, New York, USA, p 161

Bowman DA, Hodges LF (1997) An evaluation of techniques for grabbing and manipulating remote objects in immersive virtual environments. In: Proceedings of the 1997 symposium on Interactive 3D graphics - SI3D '97

Bowman DA, Koller D, Hodges LF (1997) Travel in immersive virtual environments: an evaluation of viewpoint motion control techniques. In: Proceedings - Virtual Reality Annual International Symposium. IEEE Comput. Soc. Press, pp 45–52

Bowman DA, Wingrave C, Campbell J, Ly V (2001) Using pinch gloves (TM) for both natural and abstract interaction techniques in virtual environments

Brasier E, Chapuis O, Ferey N, et al (2020) ARPads: Mid-air Indirect Input for Augmented Reality. In: 2020 IEEE International Symposium on Mixed and Augmented Reality (ISMAR). IEEE, pp 332–343

Browning G, Teather RJ (2014) Screen scaling: effects of screen scale on moving target selection. In: CHI '14 Extended Abstracts on Human Factors in Computing Systems. ACM, New York, NY, USA, pp 2053–2058

Brucker B, Brömme R, Ehrmann A, et al (2021) Touching digital objects directly on multi-touch devices fosters learning about visual contents. Comput Human Behav 119:106708. https://doi.org/10.1016/j.chb.2021.106708

Cashion J, Wingrave C, Laviola JJ (2013) Optimal 3D selection technique assignment using real-time contextual analysis. In: IEEE Symposium on 3D User Interface 2013, 3DUI 2013 - Proceedings. pp 107–110

Cashion J, Wingrave C, Laviola Jr. JJ (2012) Dense and dynamic 3D selection for game-based virtual environments. IEEE Trans Vis Comput Graph 18:634–642. https://doi.org/10.1109/TVCG.2012.40

Charness N, Holley P, Feddon J, Jastrzembski T (2004) Light Pen Use and Practice Minimize Age and Hand Performance Differences in Pointing Tasks. Hum Factors J Hum Factors Ergon Soc 46:373–384. https://doi.org/10.1518/hfes.46.3.373.50396

Clergeaud D, Guitton P (2017) Pano: Design and evaluation of a 360° through-the-lens technique. In: 2017 IEEE Symposium on 3D User Interfaces (3DUI). IEEE, pp 2–11

Cockburn A, Quinn P, Gutwin C, et al (2011) Air pointing: Design and evaluation of spatial target acquisition with and without visual feedback. Int J Hum Comput Stud 69:401–414. https://doi.org/10.1016/j.ijhcs.2011.02.005

Cohen J (1988) Statistical power analysis for the behavioural science, 2nd edn. Hillsdale, NJ

De Haan G, Koutek M, Post FH (2005) IntenSelect: Using dynamic object rating for assisting 3D object selection. In: 9th International Workshop on Immersive Projection Technology - 11th Eurographics Symposium on Virtual Environments, IPT/EGVE 2005. pp 201–209

Evangelista Belo JM, Feit AM, Feuchtner T, Grønbæk K (2021) XRgonomics: Facilitating the Creation of Ergonomic 3D Interfaces. In: Proceedings of the 2021 CHI Conference on Human Factors in Computing Systems. ACM, New York, NY, USA, pp 1–11

Facebook (2021) Best Practices. In: Oculus Dev. https://developer.oculus.com/learn/hands-design-bp/

Fitts PM (1954) The information capacity of the human motor system in controlling the amplitude of movement. J Exp Psychol 47:381–391. https://doi.org/10.1037/h0055392

Forlines C, Wigdor D, Shen C, Balakrishnan R (2007) Direct-touch vs. mouse input for tabletop displays. In: Proceedings of the SIGCHI Conference on Human Factors in Computing Systems - CHI '07. ACM Press, New York, New York, USA, pp 647–656

Forsberg A, Herndon K, Zeleznik R (1996) Aperture based selection for immersive virtual environments. In: Proceedings of the 9th annual ACM symposium on User interface software and technology - UIST '96. ACM Press, New York, New York, USA, pp 95–96

Frees S, Kessler GD, Kay E (2007) PRISM interaction for enhancing control in immersive virtual environments. ACM Trans Comput Interact 14:2. https://doi.org/10.1145/1229855.1229857

Gallo L, Ciampi M, Minutolo A (2010) Smoothed Pointing: A User-Friendly Technique for Precision Enhanced Remote Pointing. In: 2010 International Conference on Complex, Intelligent and Software Intensive Systems. IEEE, pp 712–717

Grossman T, Balakrishnan R (2006) The design and evaluation of selection techniques for 3D volumetric displays. In: Proceedings of the 19th annual ACM symposium on User interface software and technology - UIST '06. ACM Press, New York, New York, USA, p 3

Grossman T, Balakrishnan R (2005) The bubble cursor: Enhancing target acquisition by dynamic of the cursor's activation area. In: Proceedings of the SIGCHI conference on Human factors in computing systems - CHI '05. ACM Press, New York, New York, USA, p 281

Han S, Liu B, Cabezas R, et al (2020) MEgATrack: monochrome egocentric articulated hand-tracking for virtual reality. ACM Trans Graph 39:. https://doi.org/10.1145/3386569.3392452

Hart SG, Staveland LE (1988) Development of NASA-TLX (Task Load Index): Results of Empirical and Theoretical Research. Adv Psychol. https://doi.org/10.1016/S0166-4115(08)62386-9

Herndon KP, Van Dam A, Gleicher M (1994) The challenges of 3D interaction. ACM SIGCHI Bull. https://doi.org/10.1145/191642.191652

Hincapié-Ramos JD, Guo X, Moghadasian P, Irani P (2014) Consumed endurance: a metric to quantify arm fatigue of mid-air interactions. In: Proceedings of the SIGCHI Conference on Human Factors in Computing Systems. ACM, New York, NY, USA, pp 1063–1072

Hourcade JP, Bullock-Rest N (2012) How small can you go?: analyzing the effect of visual angle in pointing tasks. In: Proceedings of the 2012 ACM annual conference on Human Factors in Computing Systems - CHI '12. ACM Press, New York, New York, USA, p 213

Hwang F, Hollinworth N, Williams N (2013) Effects of Target Expansion on Selection Performance in Older Computer Users. ACM Trans Access Comput 5:1–26. https://doi.org/10.1145/2514848

ISO (2002) Ergonomic requirements for office work with visual display terminals (VDTs) — Part 9: Requirements for non-keyboard input devices (ISO 9241-9). Iso 2000 2000:54

Jang S, Stuerzlinger W, Ambike S, Ramani K (2017) Modeling Cumulative Arm Fatigue in Mid-Air Interaction based on Perceived Exertion and Kinetics of Arm Motion. In: Proceedings of the 2017 CHI Conference on Human Factors in Computing Systems - CHI '17. ACM Press, New York, New York, USA, pp 3328–3339

Jochems N, Vetter S, Schlick C (2013) A comparative study of information input devices for aging computer users. Behav Inf Technol 32:902–919. https://doi.org/10.1080/0144929X.2012.692100

Jota R, Nacenta MA, Jorge JA, et al (2010) A comparison of ray pointing techniques for very large displays. In: Proceedings - Graphics Interface

Kia K, Hwang J, Kim I-S, et al (2021) The effects of target size and error rate on the cognitive demand and stress during augmented reality interactions. Appl Ergon 97:103502. https://doi.org/10.1016/j.apergo.2021.103502

Kia K, Ishak H, Hwang J, Kim J (2020) The Effects of Target Sizes on Biomechanical Exposures and Perceived Workload during Virtual and Augmented Reality Interactions. Proc Hum Factors Ergon Soc Annu Meet 64:2107–2107. https://doi.org/10.1177/1071181320641510

Kim JH, Ari H, Madasu C, Hwang J (2020) Evaluation of the biomechanical stress in the neck and shoulders during augmented reality interactions. Appl Ergon 88:103175. https://doi.org/10.1016/j.apergo.2020.103175

Kin K, Agrawala M, DeRose T (2009) Determining the benefits of direct-touch, bimanual, and multifinger input on a multitouch workstation. In: Proceedings of Graphics Interface 2009 (GI '09). Canadian Information Processing Society, pp 119–124

König WA, Gerken J, Dierdorf S, Reiterer H (2009) Adaptive Pointing – Design and Evaluation of a Precision Enhancing Technique for Absolute Pointing Devices. In: Lecture Notes in Computer Science (including subseries Lecture Notes in Artificial Intelligence and Lecture Notes in Bioinformatics). pp 658–671

Kopper R, Bacim F, Bowman DA (2011) Rapid and accurate 3D selection by progressive refinement. In: 3DUI 2011 - IEEE Symposium on 3D User Interfaces 2011, Proceedings

Kopper R, Bowman DA, Silva MG, McMahan RP (2010) A human motor behavior model for distal pointing tasks. Int J Hum Comput Stud 68:603–615. https://doi.org/10.1016/j.ijhcs.2010.05.001

LaViola Jr. JJ, Kruijff E, McMahan RP, et al (2017) 3D user interfaces: Theory and practice. 3D User Interfaces Theory Pract

Lee S, Seo J, Kim GJ, Park C-M (2003) Evaluation of pointing techniques for ray casting selection in virtual environments. In: Pan Z, Shi J (eds) Third International Conference on Virtual Reality and Its Application in Industry. pp 38–44

Lee Y-H, Wu S-K, Liu Y-P (2013) Performance of remote target pointing hand movements in a 3D environment. Hum Mov Sci 32:511–526. https://doi.org/10.1016/j.humov.2012.02.001

Lemmerman DK, LaViola Jr. JJ (2007) Effects of interaction-display offset on user performance in surround screen virtual environments. In: Proceedings - IEEE Virtual Reality. pp 303–304

Li J, Cho I, Wartell Z (2018) Evaluation of Cursor Offset on 3D Selection in VR. In: Proceedings of the Symposium on Spatial User Interaction. ACM, New York, NY, USA, pp 120–129

Li N, Zhang Z, Liu C, et al (2021) vMirror: Enhancing the Interaction with Occluded or Distant Objects in VR with Virtual Mirrors. In: Proceedings of the 2021 CHI Conference on Human Factors in Computing Systems. ACM, New York, NY, USA, pp 1–11

Liang J, Green M (1994) JDCAD: A highly interactive 3D modeling system. Comput Graph 18:499–506. https://doi.org/10.1016/0097-8493(94)90062-0

Lin J, Harris-Adamson C, Rempel D (2019) The Design of Hand Gestures for Selecting Virtual Objects. Int J Human–Computer Interact 35:1729–1735. https://doi.org/10.1080/10447318.2019.1571783

Lindeman RW, Sibert JL, Hahn JK (1999) Hand-held windows: towards effective 2D interaction in immersive virtual environments. In: Proceedings - Virtual Reality Annual International Symposium. pp 205–212

List C, Kipp M (2019) Is Bigger Better? A Fitts' Law Study on the Impact of Display Size on Touch Performance. In: Lamas D, Loizides F, Nacke L, et al. (eds) Human-Computer Interaction – INTERACT 2019. Lecture Notes in Computer Science, vol 11748. Springer, Cham, pp 669–678

Liu Z, Vogel D, Wallace JR (2018) Applying the Cumulative Fatigue Model to Interaction on Large, Multi-Touch Displays. In: Proceedings of the 7th ACM International Symposium on Pervasive Displays. ACM, New York, NY, USA, pp 1–9

Lou X, Peng R, Hansen P, Li XA (2018) Effects of User's Hand Orientation and Spatial Movements on Free Hand

Interactions with Large Displays. Int J Human–Computer Interact 34:519–532. https://doi.org/10.1080/10447318.2017.1370811

Lu Y, Yu C, Shi Y (2020) Investigating Bubble Mechanism for Ray-Casting to Improve 3D Target Acquisition in Virtual Reality. In: 2020 IEEE Conference on Virtual Reality and 3D User Interfaces (VR). IEEE, pp 35–43

MacKenzie CL, Marteniuk RG, Dugas C, et al (1987) Three-dimensional Movement Trajectories in Fitts' task: Implications for control. Q J Exp Psychol Sect A 39:629–647. https://doi.org/10.1080/14640748708401806

MacKenzie IS (1992) Fitts' Law as a Research and Design Tool in Human-Computer Interaction. Human–Computer Interact. https://doi.org/10.1207/s15327051hci0701_3

MacKenzie IS (2015) Fitts' Throughput and the Remarkable Case of Touch-Based Target Selection. In: M K (ed) Human-Computer Interaction: Interaction Technologies. HCI 2015. Lecture Notes in Computer Science, vol 9170. Springer, Cham, pp 238–249

MacKenzie IS, Isokoski P (2008) Fitts' throughput and the speed-accuracy tradeoff. In: Proceeding of the twenty-sixth annual CHI conference on Human factors in computing systems - CHI '08. ACM Press, New York, New York, USA, p 1633

McGuffin MJ, Balakrishnan R (2005) Fitts' law and expanding targets: Experimental studies and designs for user interfaces. ACM Trans Comput Interact 12:388–422. https://doi.org/10.1145/1121112.1121115

McLaughlin AC, Rogers WA, Fisk AD (2009) Using direct and indirect input devices: Attention demands and age-related differences. ACM Trans Comput Interact 16:1–15. https://doi.org/10.1145/1502800.1502802

Mine MR (1995) Virtual environment interaction techniques. UNC Chapel Hill Comput Sci Tech Rep … 1–18. https://doi.org/10.1.1.38.1750

Mine MR, Brooks, Jr. FP, Sequin CH (1997) Moving Objects in Space: Exploiting Proprioception In Virtual-Environment Interaction. Proc ACM SIGGRAPH '97 19–26. https://doi.org/10.1145/258734.258747

Moore AG, Hatch JG, Kuehl S, McMahan RP (2018) VOTE: A ray-casting study of vote-oriented technique enhancements. Int J Hum Comput Stud 120:36–48. https://doi.org/10.1016/j.ijhcs.2018.07.003

Murata A, Iwase H (2005) Usability of Touch-Panel Interfaces for Older Adults. Hum Factors J Hum Factors Ergon Soc 47:767–776. https://doi.org/10.1518/001872005775570952

Natapov D, MacKenzie IS (2010) The trackball controller: Improving the analog stick. In: Future Play 2010: Research, Play, Share - International Academic Conference on the Future of Game Design and Technology. ACM Press, New York, New York, USA, pp 175–182

Nunnari F, Bachynskyi M, Heloir A (2016) Introducing postural variability improves the distribution of muscular loads during mid-air gestural interaction. In: Proceedings of the 9th International Conference on Motion in Games. ACM, New York, NY, USA, pp 155–160

Ortega M (2013) Hook: Heuristics for selecting 3D moving objects in dense target environments. In: IEEE Symposium on 3D User Interface 2013, 3DUI 2013 - Proceedings. pp 119–122

Penumudi SA, Kuppam VA, Kim JH, Hwang J (2020) The effects of target location on musculoskeletal load, task performance, and subjective discomfort during virtual reality interactions. Appl Ergon 84:103010. https://doi.org/10.1016/j.apergo.2019.103010

Perea P, Morand D, Nigay L (2020) Target Expansion in Context: the Case of Menu in Handheld Augmented Reality. In: Proceedings of the International Conference on Advanced Visual Interfaces. ACM, New York, NY, USA, pp 1–9

Pierce JS, Forsberg AS, Conway MJ, et al (1997) Image Plane Interaction Techniques in 3D Immersive Environments. In: Proceedings of the 1997 Symposium on Interactive 3D Graphics. p 39--ff.

Pierce JS, Stearns BC, Pausch R (1999) Voodoo Dolls: Seamless interaction at multiple scales in virtual environments. In: Proceedings of the 1999 symposium on Interactive 3D graphics - SI3D '99. ACM Press, New York, New York, USA, pp 141–145

Pohl H, Lilija K, McIntosh J, Hornbæk K (2021) Poros: Configurable Proxies for Distant Interactions in VR. In: Proceedings of the 2021 CHI Conference on Human Factors in Computing Systems. ACM, New York, NY, USA,

pp 1–12

Poupyrev I, Ichikawa T, Weghorst S, Billinghurst M (1998) Egocentric Object Manipulation in Virtual Environments: Empirical Evaluation of Interaction Techniques. Comput Graph Forum 17:41–52. https://doi.org/10.1111/1467-8659.00252

Ramcharitar A, Teather R (2018) EZCursorVR: 2D selection with virtual reality head-mounted displays. In: Proceedings - Graphics Interface. pp 114–121

Reed CL, Grubb JD, Steele C (2006) Hands up: Attentional prioritization of space near the hand. J Exp Psychol Hum Percept Perform 32:166–177. https://doi.org/10.1037/0096-1523.32.1.166

Sears A, Shneiderman B (1991) High precision touchscreens: design strategies and comparisons with a mouse. Int J Man Mach Stud 34:593–613. https://doi.org/10.1016/0020-7373(91)90037-8

Shin G, Zhu X (2011) User discomfort, work posture and muscle activity while using a touchscreen in a desktop PC setting. Ergonomics 54:733–744. https://doi.org/10.1080/00140139.2011.592604

Soukoreff RW, MacKenzie IS (2004) Towards a standard for pointing device evaluation, perspectives on 27 years of Fitts' law research in HCI. Int J Hum Comput Stud 61:751–789. https://doi.org/10.1016/j.ijhcs.2004.09.001

Steed A (2006) Towards a general model for selection in virtual environments. In: 3DUI 2006: IEEE Symposium on 3D User Interfaces 2006 - Proceedings. IEEE, pp 103–110

Steed A, Parker C (2004) 3d selection strategies for head tracked and non-head tracked operation of spatially immersive displays. In: 8th International Immersive Projection Technology Workshop

Steinicke F, Ropinski T, Hinrichs K (2006) Object Selection in Virtual Environments Using an Improved Virtual Pointer Metaphor. In: Computer Vision and Graphics. Kluwer Academic Publishers, Dordrecht, pp 320–326

Stephenson ML, Ostrander AG, Norasi H, Dorneich MC (2020) Shoulder Muscular Fatigue From Static Posture Concurrently Reduces Cognitive Attentional Resources. Hum Factors J Hum Factors Ergon Soc 62:589–602. https://doi.org/10.1177/0018720819852509

Stoakley R, Conway MJ, Pausch R (1995) Virtual reality on a WIM: interactive worlds in miniature. In: Conference on Human Factors in Computing Systems - Proceedings. pp 265–272

Stoev SL, Schmalstieg D (2002) Application and taxonomy of through-the-lens techniques. In: Proceedings of the ACM symposium on Virtual reality software and technology - VRST '02. ACM Press, New York, New York, USA, p 57

Surale HB, Gupta A, Hancock M, Vogel D (2019) TabletInVR: Exploring the Design Space for Using a Multi-Touch Tablet in Virtual Reality. In: Proceedings of the 2019 CHI Conference on Human Factors in Computing Systems. ACM, New York, NY, USA, pp 1–13

Tao D, Diao X, Wang T, et al (2021) Freehand interaction with large displays: Effects of body posture, interaction distance and target size on task performance, perceived usability and workload. Appl Ergon 93:103370. https://doi.org/10.1016/j.apergo.2021.103370

Teather RJ, Stuerzlinger W (2013) Pointing at 3d target projections with one-eyed and stereo cursors. In: Proceedings of the SIGCHI Conference on Human Factors in Computing Systems - CHI '13

Teather RJ, Stuerzlinger W (2011) Pointing at 3D targets in a stereo head-tracked virtual environment. In: 3DUI 2011 - IEEE Symposium on 3D User Interfaces 2011, Proceedings

Tseng P, Bridgeman B (2011) Improved change detection with nearby hands. Exp Brain Res 209:257–269. https://doi.org/10.1007/s00221-011-2544-z

Vanacken L, Grossman T, Coninx K (2007) Exploring the Effects of Environment Density and Target Visibility on Object Selection in 3D Virtual Environments. In: 2007 IEEE Symposium on 3D User Interfaces. IEEE

Vogel D, Balakrishnan R (2005) Distant freehand pointing and clicking on very large, high resolution displays. In: Proceedings of the 18th annual ACM symposium on User interface software and technology - UIST '05. ACM Press, New York, New York, USA, p 33

Wang Y, Mackenzie CL (1999) Effects of Orientation Disparity Between Haptic and Graphic Displays of Objects in Virtual Environments. In: INTERACT. IOS Press, Amsterdam, The Netherlands, pp 391--398

Wang Y, Yu C, Qin Y, et al (2013) Exploring the effect of display size on pointing performance. In: Proceedings of the 2013 ACM international conference on Interactive tabletops and surfaces. ACM, New York, NY, USA, pp 389–392

Ware C, Lowther K (1997) Selection Using a One-Eyed Cursor in a Fish Tank VR Environment. ACM Trans Comput Interact 4:309–322. https://doi.org/10.1145/267135.267136

Watson D, Hancock M, Mandryk RL, Birk M (2013) Deconstructing the touch experience. In: Proceedings of the 2013 ACM international conference on Interactive tabletops and surfaces. ACM, New York, NY, USA, pp 199–208

Weise M, Zender R, Lucke U (2020) How Can I Grab That?: Solving Issues of Interaction in VR by Choosing Suitable Selection and Manipulation Techniques. i-com 19:67–85. https://doi.org/10.1515/icom-2020-0011

Wickens C (1980) The Structure of Attentional Resources. In: Nickerson RS (ed) Attention and Performance Viii. Psychology Press, New York, NY, USA

Wingrave CA, Bowman DA, Ramakrishnan N (2002) Towards preferences in virtual environment interfaces. Proc 8th EGVE 63–72

Wingrave CA, Haciahmetoglu Y, Bowman DA (2006) Overcoming World in Miniature Limitations by a Scaled and Scrolling WIM. In: 3D User Interfaces (3DUI'06). IEEE, pp 11–16

Wolf D, Gugenheimer J, Combosch M, Rukzio E (2020) Understanding the Heisenberg Effect of Spatial Interaction: A Selection Induced Error for Spatially Tracked Input Devices. In: Conference on Human Factors in Computing Systems - Proceedings. ACM, New York, NY, USA, pp 1–10

Xu C, Zhou M, Zhang D, et al (2013) Guidance rays: 3D object selection based on multi-ray in dense scenario. In: Proceedings of the 12th ACM SIGGRAPH International Conference on Virtual-Reality Continuum and Its Applications in Industry - VRCAI '13. ACM Press, New York, New York, USA, pp 91–100

Yang X-D, Grossman T, Irani P, Fitzmaurice G (2011) TouchCuts and TouchZoom: enhanced target selection for touch displays using finger proximity sensing. In: Proceedings of the 2011 annual conference on Human factors in computing systems - CHI '11. ACM Press, New York, New York, USA, p 2585

Yang Z, Chen C, Lin Y, et al (2019) Effect of spatial enhancement technology on input through the keyboard in virtual reality environment. Appl Ergon 78:164–175. https://doi.org/10.1016/j.apergo.2019.03.006

Zhou X, Qin H, Xiao W, et al (2020) A Comparative Usability Study of Bare Hand Three-Dimensional Object Selection Techniques in Virtual Environment. Symmetry (Basel) 12:1723. https://doi.org/10.3390/sym12101723

# Appendix. Descriptive statistics and ANOVA results

**Table 1.** F, p-values, and $\eta^2$ from ANOVA results of the effects of technique, target size, and target distance on performance, perceived workload, and user behavior measures

| | Measures | Values | Source of variation | | | | | | |
|---|---|---|---|---|---|---|---|---|---|
| | | | TQ | TS | TD | TQ×TS | TQ×TD | TS×TD | TQ×TS×TD |
| Performance | Movement time (s) | $F_{2,689}$ | 317.38 | 648.05 | 114.82 | 43.12 | 1.65 | 2.87 | 2.53 |
| | | $p$ | ***<0.001 | ***<0.001 | ***<0.001 | ***<0.001 | 0.193 | 0.091 | 0.081 |
| | | $\eta^2$ | 0.243 | 0.249 | 0.044 | 0.033 | 0.001 | 0.001 | 0.002 |
| | Error rate (%) | $F_{2,689}$ | 34.20 | 332.48 | 55.95 | 20.22 | 3.37 | 25.62 | 4.19 |
| | | $p$ | ***<0.001 | ***<0.001 | ***<0.001 | ***<0.001 | *0.035 | ***<0.001 | *0.016 |
| | | $\eta^2$ | 0.048 | 0.232 | 0.039 | 0.028 | 0.005 | 0.018 | 0.006 |
| | Throughput (bit/s) | $F_{2,689}$ | 451.07 | 150.78 | 3.16 | 5.59 | 3.50 | 10.07 | 1.67 |
| | | $p$ | ***<0.001 | ***<0.001 | 0.076 | **0.004 | *0.031 | **0.002 | 0.190 |
| | | $\eta^2$ | 0.408 | 0.068 | 0.001 | 0.005 | 0.003 | 0.005 | 0.002 |
| Perceived workload | Mental demand | $F_{2,209}$ | 60.74 | 226.25 | 27.34 | 4.82 | 1.55 | 3.04 | 2.40 |
| | | $p$ | ***<0.001 | ***<0.001 | ***<0.001 | **0.009 | 0.215 | 0.083 | 0.093 |
| | | $\eta^2$ | 0.153 | 0.285 | 0.034 | 0.012 | 0.004 | 0.004 | 0.006 |
| | Physical demand | $F_{2,209}$ | 39.96 | 198.59 | 32.77 | 3.49 | 0.32 | 6.34 | 1.52 |
| | | $p$ | ***<0.001 | ***<0.001 | ***<0.001 | *0.032 | 0.728 | *0.013 | 0.220 |
| | | $\eta^2$ | 0.112 | 0.277 | 0.046 | 0.010 | 0.001 | 0.009 | 0.004 |
| | Temporal demand | $F_{2,209}$ | 39.76 | 144.71 | 17.67 | 3.95 | 0.06 | 6.49 | 2.83 |
| | | $p$ | ***<0.001 | ***<0.001 | ***<0.001 | *0.021 | 0.945 | *0.012 | 0.062 |
| | | $\eta^2$ | 0.105 | 0.192 | 0.023 | 0.010 | <0.001 | 0.009 | 0.007 |
| | Performance | $F_{2,209}$ | 21.43 | 181.13 | 24.53 | 8.56 | 1.22 | 5.28 | 2.91 |
| | | $p$ | ***<0.001 | ***<0.001 | ***<0.001 | ***<0.001 | 0.296 | *0.023 | 0.057 |
| | | $\eta^2$ | 0.072 | 0.303 | 0.041 | 0.029 | 0.004 | 0.009 | 0.010 |
| | Effort | $F_{2,209}$ | 52.15 | 231.50 | 25.03 | 2.15 | 0.25 | 6.62 | 2.75 |
| | | $p$ | ***<0.001 | ***<0.001 | ***<0.001 | 0.119 | 0.783 | *0.011 | 0.066 |
| | | $\eta^2$ | 0.127 | 0.283 | 0.031 | 0.005 | 0.001 | 0.008 | 0.007 |
| | Frustration | $F_{2,209}$ | 30.29 | 166.82 | 19.42 | 4.90 | 0.16 | 5.87 | 2.19 |
| | | $p$ | ***<0.001 | ***<0.001 | ***<0.001 | *0.008 | 0.848 | *0.016 | 0.114 |
| | | $\eta^2$ | 0.084 | 0.232 | 0.027 | 0.014 | <0.001 | 0.008 | 0.006 |
| | Weighted rating | $F_{2,209}$ | 55.38 | 276.92 | 33.70 | 5.91 | 0.67 | 6.92 | 3.63 |
| | | $p$ | ***<0.001 | ***<0.001 | ***<0.001 | **0.003 | 0.511 | **0.009 | *0.028 |
| | | $\eta^2$ | 0.127 | 0.318 | 0.039 | 0.014 | 0.002 | 0.008 | 0.008 |
| User behavior | Hand movement (m) | $F_{2,689}$ | 112.66 | 34.51 | 43.66 | 0.83 | 5.18 | 2.68 | 1.02 |
| | | $p$ | ***<0.001 | ***<0.001 | ***<0.001 | 0.439 | **0.006 | 0.102 | 0.361 |
| | | $\eta^2$ | 0.207 | 0.032 | 0.040 | 0.002 | 0.010 | 0.002 | 0.002 |
| | Head movement (m) | $F_{2,689}$ | 79.87 | 521.47 | 285.28 | 22.49 | 23.13 | 28.81 | 9.70 |
| | | $p$ | ***<0.001 | ***<0.001 | ***<0.001 | ***<0.001 | ***<0.001 | ***<0.001 | ***<0.001 |
| | | $\eta^2$ | 0.059 | 0.192 | 0.105 | 0.017 | 0.017 | 0.011 | 0.007 |
| | Actual target depth | $F_{2,689}$ | 334002.29 | 0.03 | 4.98 | 1.92 | 1.73 | 0.45 | 0.49 |
| | | $p$ | ***<0.001 | 0.852 | *0.026 | 0.148 | 0.178 | 0.503 | 0.612 |
| | | $\eta^2$ | 0.998 | <0.001 | <0.001 | <0.001 | <0.001 | <0.001 | <0.001 |
| | Adjusted target visual angle (°) | $F_{2,689}$ | 290.75 | 1548.34 | 176.88 | 11.16 | 46.14 | 14.31 | 4.33 |
| | | $p$ | ***<0.001 | ***<0.001 | ***<0.001 | ***<0.001 | ***<0.001 | ***<0.001 | *0.014 |
| | | $\eta^2$ | 0.171 | 0.454 | 0.052 | 0.007 | 0.027 | 0.004 | 0.003 |

\* TQ=Technique, TS=Target size, TD=Target distance; \*$p < 0.05$, \*\*$p < 0.01$, \*\*\*$p < 0.001$

**Table 2.** Mean (M) and standard deviation (SD) of performance, perceived workload, and user behavior measures by technique

| Measures | | Technique | | | | | |
|---|---|---|---|---|---|---|---|
| | | Raycasting | | ViewfinderVR-Ray | | ViewfinderVR-Touch | |
| | | M | SD | M | SD | M | SD |
| Performance | Movement time (s) | 2.38 | 1.13 | 1.69 | 0.60 | 1.29 | 0.52 |
| | Error rate (%) | 17.08 | 19.28 | 9.30 | 12.42 | 10.36 | 13.79 |
| | Throughput (bit/s) | 1.64 | 0.61 | 2.31 | 0.62 | 3.16 | 0.97 |
| Perceived workload | Mental demand | 57.06 | 25.64 | 43.31 | 25.57 | 30.94 | 24.57 |
| | Physical demand | 57.13 | 26.17 | 42.69 | 24.39 | 35.63 | 25.60 |
| | Temporal demand | 56.13 | 26.27 | 42.88 | 25.91 | 34.31 | 26.79 |
| | Performance | 50.81 | 28.92 | 40.31 | 26.52 | 32.25 | 27.12 |
| | Effort | 62.31 | 26.03 | 45.13 | 27.65 | 37.63 | 27.90 |
| | Frustration | 52.25 | 28.57 | 37.75 | 27.75 | 32.06 | 28.21 |
| | Weighted rating | 57.05 | 25.52 | 43.22 | 25.07 | 33.71 | 25.26 |
| User behavior | Hand movement (m) | 1.61 | 1.34 | 1.77 | 1.62 | 3.64 | 2.33 |
| | Head movement (m) | 0.36 | 0.24 | 0.34 | 0.24 | 0.50 | 0.32 |
| | Actual target depth (m) | 5.00 | 0.09 | 0.47 | 0.11 | 0.40 | 0.06 |
| | Adjusted target visual angle (°) | 2.00 | 1.00 | 3.37 | 1.91 | 3.68 | 1.80 |

**Table 3.** Mean (M) and standard deviation (SD) of performance, perceived workload, and user behavior measures by technique, target size, and target distance

| Measures | | TQ | Raycasting | | | | ViewfinderVR-Ray | | | | ViewfinderVR-Touch | | | |
|---|---|---|---|---|---|---|---|---|---|---|---|---|---|---|
| | | TD | 1m | | 2m | | 1m | | 2m | | 1m | | 2m | |
| | | TS | M | SD | M | SD | M | SD | M | SD | M | SD | M | SD |
| Performance | Movement time (s) | 3° | 1.45 | 0.35 | 1.94 | 0.95 | 1.16 | 0.30 | 1.49 | 0.37 | 0.90 | 0.22 | 1.04 | 0.26 |
| | | 1° | 2.87 | 1.06 | 3.28 | 0.95 | 1.81 | 0.49 | 2.29 | 0.52 | 1.39 | 0.42 | 1.83 | 0.52 |
| | Error rate (%) | 3° | 4.70 | 9.53 | 7.44 | 9.22 | 4.09 | 7.19 | 5.40 | 7.02 | 2.12 | 5.12 | 4.12 | 7.06 |
| | | 1° | 25.48 | 19.44 | 30.72 | 20.91 | 8.94 | 12.78 | 18.75 | 15.03 | 9.44 | 10.30 | 25.76 | 15.41 |
| | Throughput (bit/s) | 3° | 1.86 | 0.57 | 1.98 | 0.69 | 2.47 | 0.67 | 2.47 | 0.62 | 3.46 | 1.01 | 3.52 | 0.87 |
| | | 1° | 1.34 | 0.50 | 1.39 | 0.37 | 2.29 | 0.60 | 2.00 | 0.48 | 3.03 | 0.98 | 2.65 | 0.74 |
| Perceived workload | Mental demand | 3° | 35.50 | 19.05 | 41.50 | 20.80 | 27.50 | 21.96 | 36.50 | 20.57 | 14.50 | 12.43 | 19.75 | 17.58 |
| | | 1° | 72.25 | 17.80 | 79.00 | 11.31 | 49.75 | 23.03 | 59.50 | 25.49 | 32.75 | 20.04 | 56.75 | 22.90 |
| | Physical demand | 3° | 33.75 | 20.88 | 45.00 | 23.34 | 29.25 | 21.65 | 33.00 | 16.82 | 20.50 | 17.49 | 24.75 | 16.68 |
| | | 1° | 68.75 | 15.88 | 81.00 | 11.90 | 46.75 | 20.97 | 61.75 | 24.33 | 37.50 | 24.69 | 59.75 | 23.48 |
| | Temporal demand | 3° | 36.50 | 21.82 | 43.75 | 24.82 | 30.75 | 25.09 | 36.25 | 19.86 | 24.75 | 24.69 | 22.00 | 20.08 |
| | | 1° | 68.25 | 19.95 | 76.00 | 15.52 | 46.00 | 28.26 | 58.50 | 22.36 | 35.00 | 24.15 | 55.50 | 25.63 |
| | Performance | 3° | 27.00 | 20.97 | 32.00 | 20.21 | 24.25 | 22.45 | 37.25 | 23.70 | 16.00 | 14.40 | 16.50 | 15.16 |
| | | 1° | 68.25 | 21.42 | 76.00 | 14.58 | 42.00 | 23.66 | 57.75 | 26.39 | 34.75 | 25.85 | 61.75 | 22.18 |
| | Effort | 3° | 39.75 | 22.63 | 48.50 | 22.99 | 29.00 | 25.58 | 35.00 | 22.41 | 22.75 | 21.06 | 22.75 | 20.66 |
| | | 1° | 76.75 | 15.43 | 84.25 | 8.77 | 50.00 | 22.58 | 66.50 | 25.00 | 41.50 | 27.91 | 63.50 | 19.99 |
| | Frustration | 3° | 30.00 | 21.11 | 38.00 | 25.58 | 25.25 | 25.98 | 30.50 | 21.65 | 18.00 | 22.14 | 17.75 | 18.69 |
| | | 1° | 65.50 | 21.65 | 75.50 | 18.43 | 41.75 | 26.79 | 53.50 | 29.07 | 34.75 | 25.31 | 57.75 | 26.52 |
| | Weighted rating | 3° | 33.75 | 18.56 | 41.63 | 20.17 | 27.28 | 22.90 | 35.45 | 19.23 | 18.80 | 15.47 | 20.10 | 16.64 |
| | | 1° | 72.63 | 15.25 | 80.20 | 9.08 | 47.93 | 22.00 | 62.20 | 22.41 | 35.83 | 23.08 | 60.10 | 20.93 |
| User behavior | Hand movement (m) | 3° | 1.07 | 1.25 | 1.55 | 0.45 | 1.25 | 1.77 | 1.67 | 1.71 | 2.70 | 1.25 | 3.63 | 2.19 |
| | | 1° | 1.48 | 1.21 | 2.33 | 1.80 | 1.87 | 1.80 | 2.28 | 0.87 | 3.22 | 1.00 | 5.00 | 3.41 |
| | Head movement (m) | 3° | 0.21 | 0.11 | 0.32 | 0.21 | 0.20 | 0.14 | 0.30 | 0.19 | 0.24 | 0.12 | 0.41 | 0.20 |
| | | 1° | 0.39 | 0.26 | 0.52 | 0.25 | 0.36 | 0.23 | 0.52 | 0.28 | 0.47 | 0.21 | 0.87 | 0.32 |
| | Actual target depth (m) | 3° | 5.01 | 0.08 | 5.01 | 0.09 | 0.45 | 0.09 | 0.48 | 0.11 | 0.39 | 0.05 | 0.40 | 0.05 |
| | | 1° | 4.99 | 0.08 | 5.00 | 0.09 | 0.46 | 0.12 | 0.48 | 0.13 | 0.40 | 0.05 | 0.40 | 0.08 |
| | Adjusted target visual angle (°) | 3° | 2.99 | 0.05 | 2.99 | 0.05 | 5.55 | 2.15 | 3.75 | 0.95 | 5.67 | 1.80 | 4.34 | 1.02 |
| | | 1° | 1.00 | 0.02 | 1.00 | 0.02 | 2.56 | 0.68 | 1.64 | 0.36 | 2.77 | 0.63 | 1.96 | 0.26 |

* TQ=Technique, TS=Target size, TD=Target distance